\documentclass[%
 reprint,
 amsmath,amssymb,
 aps,
]{revtex4-2}

\usepackage{graphicx}
\usepackage{dcolumn}
\usepackage{bm}
\usepackage{gensymb} 

\begin{document}


\title{Polarization-independent
  isotropic nonlocal metasurfaces \\
  with wavelength-controlled functionality}

\author{Olivia Y. Long}
\affiliation{Department of Applied Physics, Stanford University, Stanford, California 94305, USA}
\author{Cheng Guo}
\affiliation{Department of Applied Physics, Stanford University, Stanford, California 94305, USA}
\author{Weiliang Jin}
\affiliation{Ginzton Laboratory and Department of Electrical Engineering,
Stanford University, Stanford, California 94305, USA}
\author{Shanhui Fan}
\email{shanhui@stanford.edu}
\affiliation{Department of Applied Physics, Stanford University, Stanford, California 94305, USA}
\affiliation{Ginzton Laboratory and Department of Electrical Engineering,
Stanford University, Stanford, California 94305, USA}




\date{\today}

\begin{abstract}
Flat optics has demonstrated great advances in miniaturizing conventional, bulky optical elements due to the recent developments in metasurface design. Specific applications of such designs include spatial differentiation and the compression of free space.
However, metasurfaces designed for such applications are often polarization-dependent and are designed for a single functionality. 
In this work, we introduce a polarization-independent metasurface structure by designing guided resonances with degenerate band curvatures in a photonic crystal slab. Our device can perform both free-space compression and spatial differentiation when operated at different frequencies at normal incidence. This work demonstrates the promise of dispersion engineering in metasurface design to create ultrathin devices with polarization-independent functionality.
\end{abstract}

\maketitle



\section{Introduction}

There has been significant recent interests in flat optics and metasurfaces, which allow for arbitrary manipulation of the phase and amplitude of optical wavefronts using ultrathin optical elements at sub-wavelength scales \cite{yu_capasso_flat_optics_metasurf_review_2014,yu_capasso_generalized_snell_science_2011,Silva_2014, Capasso_flat_optics_perspective_2018}.
The diverse applications of flat optics include replacing optical elements used in phone camera modules \cite{capasso_flat_optics_dispersion_eng_nature_review_2020}, optical microscopy \cite{zhou2020flat}, holograms \cite{Ni_metasurf_hologram_nature_comm_2013}, augmented reality \cite{Lee_metasurf_aug_reality_2018}, and integrated optics. \cite{Capasso_flat_optics_perspective_2018, yu_capasso_flat_optics_metasurf_review_2014, Banerji_imaging_flat_optics_2019}
%
%
%
%
%

Existing metasurfaces can be approximately classified into two types: the local metasurface \cite{capasso_flat_optics_dispersion_eng_nature_review_2020, yu_capasso_generalized_snell_science_2011, Silva_2014}, where the transmission or reflection coefficients are a function of position, and the nonlocal metasurface, where such coefficients are a function of wave vector \cite{kwon2018}. 
%
Nonlocal metasurfaces have found fruitful applications in 
optical analog computing.
The key advantages of optical analog computing include improved power efficiency, high-speed operation, and high-throughput data processing compared to its digital counterpart.
\cite{zangeneh_review_2020}.
Mathematical operations such as differentiation \cite{ zhu_plasmonic_2017, Komar_magnetic_Mie_diff_metasurf_ACS_2021, wang2020compact, Guo_phc_diff_optica_2018, kwon2020, tengfeng_topolog_diff_2021, Long_OL_21, Silva_2014, tengfeng_spinhall_2019, Zhou_PNAS_2019, he_geom_SHE_2020, cordaro_metasurf_2019, Bykov_diff_grating_18, Pan_laplace_metasurf_21} and integration \cite{Pors_diff_integ_reflect_metasurf_nanolatters_2015, Babashah_metasurf_integrator_JOSAB, Silva_2014} have been demonstrated. 
In particular, there have been realizations of the Laplacian operator \cite{zhou2020flat, Guo_phc_diff_optica_2018,Long_OL_21, kwon2020, wang2020compact, Pan_laplace_metasurf_21}, which provides isotropic, two-dimensional differentiation. This is of particular importance in image processing, where differentiation is used for edge detection \cite{gonzalez_digital_image_book, marr1980theory}. Nonlocal metasurfaces have also been used for the compression of free-space \cite{aobo_2021,  reshef_nature_2021, Guo_squeeze_optica_2020}, which is important for the miniaturization and integration of optical systems.  

For the applications mentioned above, it is desirable for the devices to process both polarizations and to have polarization-independent responses. Existing designs of nonlocal metasurfaces for Laplacian operators \cite{Guo_phc_diff_optica_2018, kwon2020} and for free-space compression \cite{Guo_squeeze_optica_2020} are based on guided resonances in photonic crystal slabs. These guided resonances, however, have strong polarization dependency, which presents a limitation in practical applications.   
 
In this work, we present a polarization-independent, photonic crystal slab design that can be used as both a space compressor and a differentiator when operated at different frequencies. Our nonlocal metasurface features the use of guided resonances with degenerate band curvatures. Moreover, our device operates in transmission mode at normal incidence, allowing for ease of integration into optical systems. 

The rest of the paper is organized as follows: we first briefly review the physics of guided mode resonances in photonic crystal slabs and how to employ such resonances for our applications. Then, we describe the design of our device in light of the necessary conditions for each application. Finally, we numerically demonstrate the multifunctionality of our structure and assess its performance.

\section{Theory}
\subsection{Operation principle of \\ space compressor and differentiator}

In this section, we briefly review the operation principle of a spatial compressor and a differentiator designed using photonic crystal slabs. 
Photonic crystal slabs can support guided resonances 
\cite{Fan_guided_res_phc_PRB_2002}, which can couple to external light and thus significantly influence the transmission and reflection spectra. 
Because we are manipulating the electric field in momentum space, we first decompose the incident, transverse scalar field $E(x,y)$ in terms of the spatial wave vector components $k_x$ and $k_y$ \cite{goodman_fourier_book}: 

\begin{equation}
    E(x,y) = \iint E(k_x, k_y) e^{ik_xx}e^{ik_yy}dk_x dk_y
\end{equation}
where $E(k_x, k_y)$ is the Fourier transform of the field $E(x,y)$. The transmission coefficient $t(\omega, \mathbf{k})$ at an operating frequency $\omega$ and as a function of the in-plane wave vector $\mathbf{k} = (k_x, k_y)$ is given by the equation: 
\cite{Fan_guided_res_phc_PRB_2002}
\begin{equation} \label{transmission_eqn}
    t(\omega, \mathbf{k}) = t_d - (t_d \pm r_d)\frac{\gamma(\mathbf{k})}{i(\omega - \omega_0(\mathbf{k})) + \gamma(\mathbf{k})}
\end{equation}
where $t_d$ and $r_d$ are the direct transmission and reflection coefficients, and $\omega_0(\mathbf{k})$ and $\gamma(\mathbf{k})$ are the resonant frequency and linewidth of a guided resonance at the wave vector $\mathbf{k}$. 
The plus or minus sign corresponds to the case where the guided resonance has even or odd symmetry with respect to the transverse plane of the slab, respectively.
The resulting transmission spectrum near a resonant frequency exhibits a Fano lineshape, which finds many applications in photonics due to its rapid variation between total transmission and reflection \cite{limonov_fano_review_2017}. 

With this knowledge of transmission through a photonic crystal slab, we will now briefly review the general conditions needed to achieve the two functionalities studied in this work: free-space compression and isotropic, second-order differentiation. For these applications, since we will mostly be focusing on the operation of the slab at a single frequency $\omega = \omega_{\text{op}}$, for the rest of the paper we will denote $t(\omega_{\text{op}}, \mathbf{k})$ as the transfer function $t(\mathbf{k})$ of our device. 

In this paper, we consider a light beam normally incident upon the slab. Thus, we are particularly interested in the response of the structure near $\mathbf{k} = 0$. Both functionalities that we consider in this work require an isotropic transfer function $t(|\mathbf{k}|)$.  Therefore, from Eq. \ref{transmission_eqn}, both $\omega_0(\mathbf{k})$ and $\gamma(\mathbf{k})$ should be isotropic, i.e. they should depend only on $|\mathbf{k}|$, which is the length of the wave vector $\mathbf{k}$. Thus, the slab should exhibit an isotropic band structure.

\begin{figure}[b]
\includegraphics[width=\linewidth]{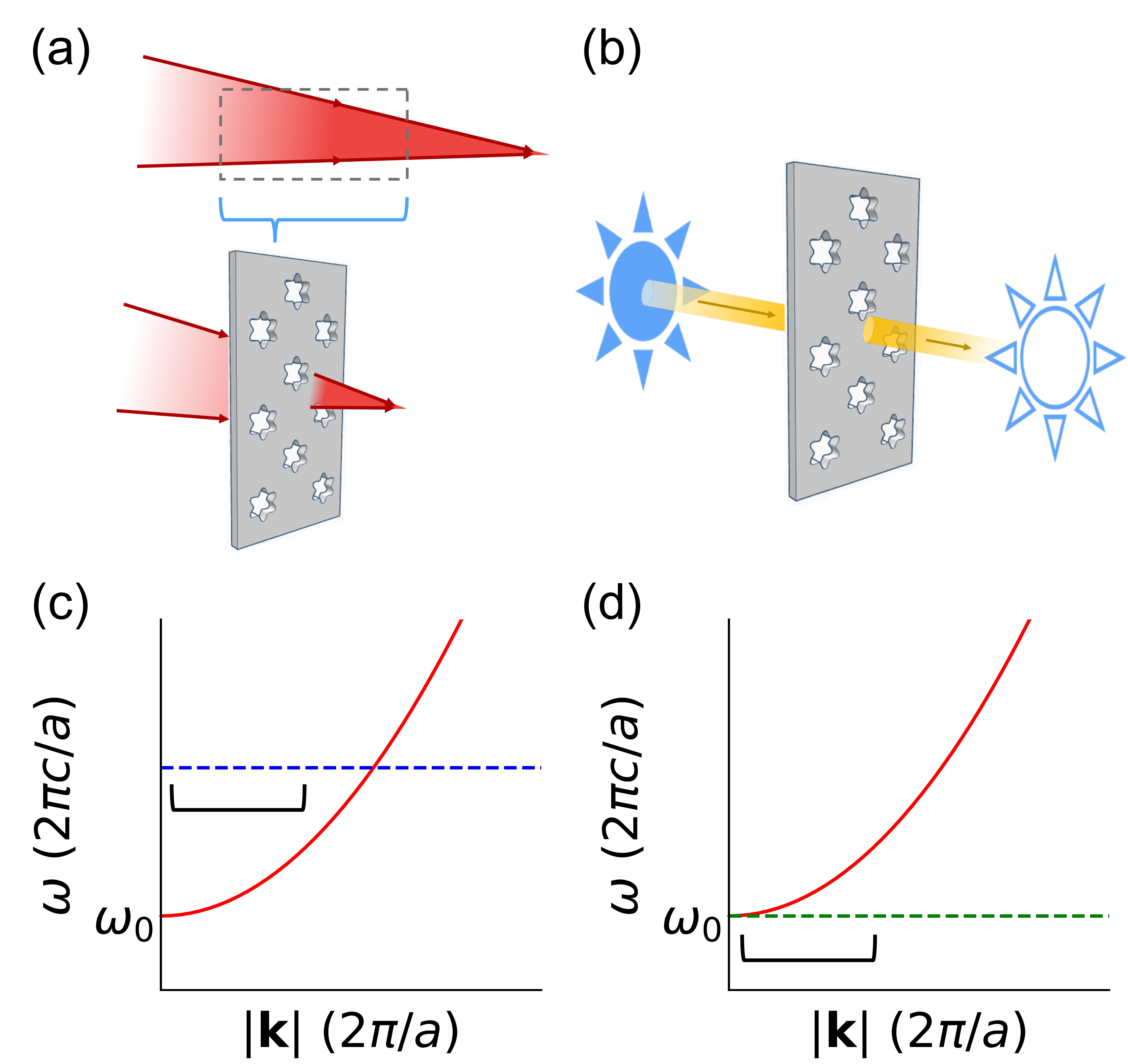}
\caption{\label{apps_schematics} Operating principle of 
(a) space compressor and (b) spatial differentiator applications using the same photonic crystal slab. Band diagram schematics showing the operating conditions of the (c) space compressor and the (d) differentiator. Operating frequencies are denoted by the dashed lines, and the black brackets denote the operating wave vector range.}
\end{figure}

An illustration of a space-compression device is shown in Figure \ref{apps_schematics}(a). This device aims to achieve the same transfer function as free-space propagation over a distance $d$, but using a device with a thickness much smaller than $d$. 
To achieve free-space compression for light incident near the direction normal to the slab, the transfer function $t(\mathbf{k})$ of the device must satisfy the following conditions \cite{Guo_squeeze_optica_2020}:
\begin{enumerate}
\item $|t( \mathbf{k})| = 1$ for $\mathbf{k}$ in the wave vector range of operation 
\item $\text{arg}[t(\mathbf{k})] \propto |\mathbf{k}|^2$ to lowest order in $\mathbf{k}$ (see Eqn. 18 in Ref. \citenum{Guo_squeeze_optica_2020})
\end{enumerate}

To satisfy the two conditions above, we choose the operating frequency $\omega_{\text{op}}$ to be near, but not exactly at, the frequencies of the guided resonance band within the wave vector range of interest (Figure \ref{apps_schematics}(c)). With this choice, the magnitude of the transfer function is not strongly affected by the presence of the guided resonance, i.e. $t(\mathbf{k}) \approx t_d(\mathbf{k})$, and we can choose the thickness of the structure such that $t_d(\mathbf{k}) \approx 1$ \cite{Fan_guided_res_phc_PRB_2002}. On the other hand, the phase of the transfer function is affected by the presence of the guided resonance, which gives rise to the $|\mathbf{k}|^2$ behavior of the phase response. 
From the curvatures of the quadratic phase dependency on the wave vector, we can calculate the effective distance $d_{\text{eff}}$, which is the total distance of replaced free space \cite{Guo_squeeze_optica_2020}. Thus, as a measure of our device performance, we define a compression ratio $d_{\text{eff}}/d$, where $d$ is the thickness of the device. 

%

Using the same device, 
we can also perform differentiation, as shown schematically in Figure \ref{apps_schematics}(b). To achieve a Laplacian operator, 
the transfer function must satisfy the following conditions  \cite{Guo_phc_diff_optica_2018}:

\begin{enumerate}
\item $t( \mathbf{k}) = 0$ at $\mathbf{k} = 0$
\item $|t( \mathbf{k})| \propto |\mathbf{k}|^2$ to lowest order in $\mathbf{k}$
\end{enumerate}

To satisfy the two conditions above, we choose $\omega_{\text{op}} = \omega_0(\mathbf{k} = 0)$ (Figure \ref{apps_schematics}(d)).  We further choose the slab parameters such that $t_d(\omega_{\text{op}}, \mathbf{k} = 0) = 1$. As derived in Ref. \citenum{Guo_phc_diff_optica_2018}, with these choices, the transmission coefficient $t( \mathbf{k})$ near $\mathbf{k} = 0$ is given by the relation:
\begin{equation}\label{eqn2}
    t(\mathbf{k}) = - \frac{i}{\gamma(0)}(\omega_0(\mathbf{k}) - \omega_0(0))
\end{equation}
Thus, when the dispersion relation $\omega_0(\mathbf{k}) \propto |\mathbf{k}|^2$, $t( \mathbf{k}) \propto |\mathbf{k}|^2$ as well \cite{Guo_phc_diff_optica_2018}. The structure then operates as a differentiator implementing a Laplacian operator. 






\subsection{Design of photonic crystal slab}
The operating principles as outlined in the previous section have been implemented using photonic crystal slab structures. To obtain the necessary isotropic band structure, one can use a photonic crystal slab with $C_{6v}$ symmetry \cite{Guo_phc_diff_optica_2018}.
Moreover, at the $\Gamma$ point where $\mathbf{k} = 0$, the eigenmode must be doubly degenerate in order to couple to external light \cite{Fan_guided_res_phc_PRB_2002}. From a group theory standpoint, the mode must match the two-dimensional, irreducible representation of the external plane waves, given by the $E_1$ representation of the $C_{6v}$ point group \cite{sakoda_hexag_phc_symm_PRB_2001, sakoda_symmetry_2d_hexag_prb_1995, dresselhaus2007group}. However, one limitation of the previous designs is the polarization dependency. Away from $\mathbf{k} = 0$, the degeneracy of the mode is typically lifted, and the two bands couple to the two polarizations in a different fashion, leading to a polarization-dependent response. Such a polarization dependency is generally undesirable for practical applications.

\begin{figure}[b]
\includegraphics[width=\linewidth]{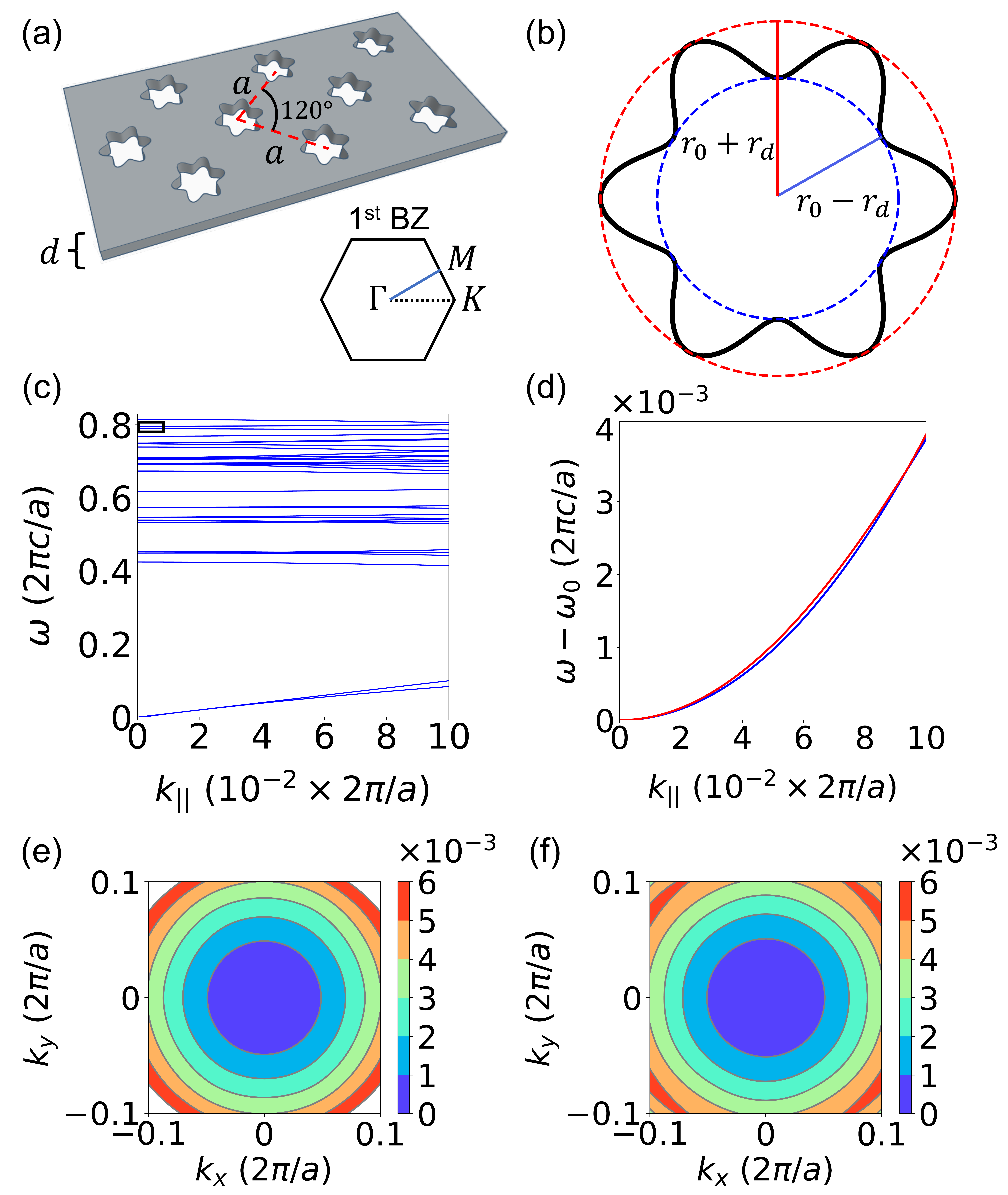}
\caption{\label{phc_slab_design} (a) Photonic crystal slab design with thickness $d=0.32a$, $\epsilon = 12$, and air holes arranged in a hexagonal lattice with lattice constant $a$. Inset: The first Brillouin zone of the lattice with high symmetry points $\Gamma$, $M$, and $K$. (b) Daisy shape of air holes described by the equation: $r(\phi) = r_0 + r_d \cos{6\phi}$. (c) Overall band structure of the photonic crystal slab design along the $\Gamma K$ direction, including both even and odd-symmetry modes. The small black box indicates where the degenerate bands lie. (d) Band structure of the doubly degenerate bands with degenerate curvatures near $\mathbf{k} = 0$ plotted along the $\Gamma K$ direction, calculated using GME. $\omega_0=0.7902 \times 2\pi c/a$ corresponds to the frequency of the degenerate eigenmodes. The red and blue colors distinguish the two bands. (e) Isofrequency contour plot of degenerate eigenmode 1 and (f) eigenmode 2 calculated using GME. }
\end{figure}

To overcome the  issue associated with polarization dependency, we modify the photonic crystal slab structures previously considered. Our photonic crystal slab consists of a hexagonal lattice of air holes in a dielectric slab with lattice constant $a$, thickness $d=0.32a$, and relative permittivity $\epsilon = 12$, as illustrated in Figure \ref{phc_slab_design}(a). 
The first Brillouin zone of the hexagonal lattice is shown in the inset of Figure \ref{phc_slab_design}(a), where the labeled high symmetry points $\Gamma$, $M$, and $K$ correspond to $(k_x, k_y) = (0, 0)$, $(k_x, k_y) = (\pi, \pi/\sqrt{3})$, and $(k_x, k_y) = (4\pi/3, 0)$, respectively. All wave vectors are in units of $2 \pi /a$.  For our purpose,
the air hole shape must respect the necessary $C_{6v}$ symmetry, as shown in Figure \ref{phc_slab_design}(b). We select the daisy shape as defined by the polar equation that relates the radius $r$ to the angle $\phi$: $r(\phi) = r_0 + r_d \cos{6\phi}$, where $r_0 = 0.21a$ and $r_d = 0.04a$. In contrast to circular air holes with one radius parameter, we now have two tunable parameters: $r_d$ and $r_0$. This additional degree of freedom turns out to be sufficient to achieve near-degenerate bands away from $\mathbf{k} = 0$. 
Such a ``daisy photonic crystal" using different parameter values has previously been shown to generate zero-index modes with no radiation loss at the $\Gamma$ point \cite{momchil_daisy_PRL_2018}.

In Figure \ref{phc_slab_design}(c), we plot the band diagram of the photonic crystal slab, including both even and odd-symmetry modes with respect to the transverse plane of the slab.
Figure \ref{phc_slab_design}(d) shows the band structure along the $\Gamma K$ direction near the frequency $\omega_0 = 0.7902 \times 2\pi c/a$, which is a resonant frequency of the structure at $\mathbf{k} = 0$. These eigenmodes are two-fold degenerate at $\mathbf{k} = 0$ and have the correct symmetry to couple to normally incident light. Away from $\mathbf{k} = 0$, the degeneracy persists, as desired. 
%
We note that at a frequency near $\omega_0$, there are only zeroth order propagating diffracted waves, since $\omega_0 \leq 1.15 \times 2\pi c/a$
\cite{kittel_intro_solid_state_text, sakoda_hexag_phc_symm_PRB_2001}.   
%
%
%
Figures \ref{phc_slab_design}(e) and \ref{phc_slab_design}(f) show the isofrequency contour plots of the two bands shown in Figure \ref{phc_slab_design}(d). Both modes show isotropic band structures around the $\Gamma$ point and thus have the required $|\mathbf{k}|^2$ dependency. The contours are also quite similar in their frequency dependency, confirming the near-degeneracy between the two bands. All band structures in Figures \ref{phc_slab_design}(c)$-$(f) show the real part of the eigenmode frequencies and were computed using the method of guided-mode expansion (GME) \cite{Minkov_legume_2020, andreani_GME_PRB_2006}.

\begin{figure*}[t]
\includegraphics[width=0.85\linewidth]{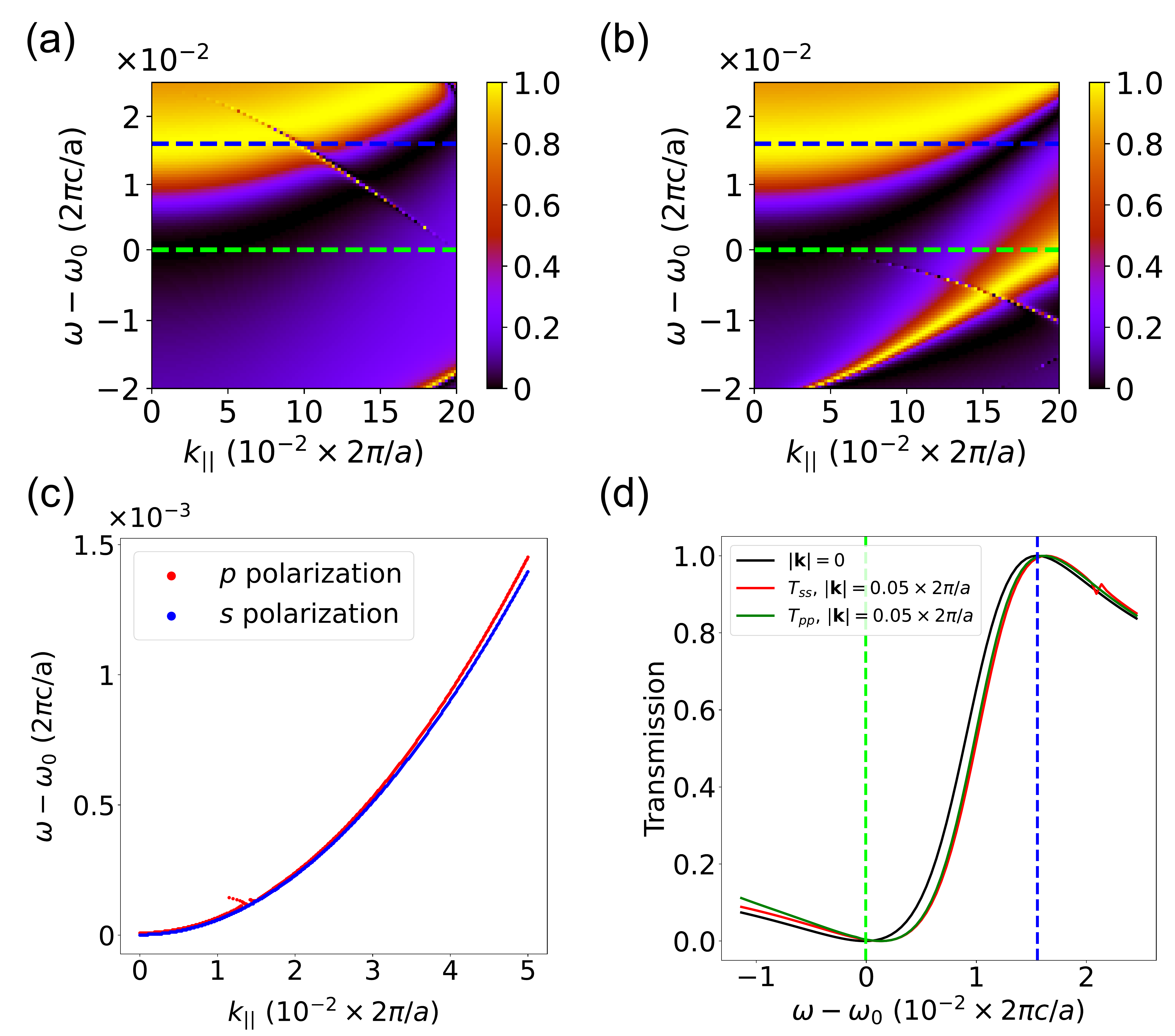}
\caption{\label{transmission_plots} (a) Transmission $|t_{ss}|^2$ of $s$-polarized light and (b) transmission $|t_{pp}|^2$ of $p$-polarized light as a function of in-plane wave vector $k_{||}$ computed along $\phi=0$ ($\Gamma K$) direction using RCWA. $\omega_0=0.7902 \times 2\pi c/a$ corresponds to the frequency of the degenerate eigenmode at normal incidence. 
(c) Transmission nodal line of $s$- and $p$-polarized light along $\phi=0$ ($\Gamma K$) direction obtained from RCWA calculations. (d) Transmission as a function of frequency for normally incident light, as well as off-normal $s$-polarized and $p$-polarized light at $|\mathbf{k}| = 0.05 \times 2\pi /a$, computed along the $\phi = 0$ direction using RCWA. In panels (a), (b), and (d), the dashed green line indicates the operating frequency for the differentiator and the blue dashed line indicates that for the free-space compressor.}
\end{figure*}

In the frequency range near $\omega_0 = 0.7902 \times 2\pi c/a$, we calculate the transmission as a function of the frequency detuning from $\omega_0$ and the in-plane wave vector $\mathbf{k}_{||}$ using the rigorous coupled-wave analysis (RCWA) method \cite{Jin_RCWA_2020, xsede}. Figures \ref{transmission_plots}(a) and \ref{transmission_plots}(b) show the transmission plots for $s$- and $p$-polarized incident light, respectively. At each wave vector $\mathbf{k}$, there is a frequency near the resonant frequency $\omega_0(\mathbf{k})$ where the transmission is 0. 
In Figure \ref{transmission_plots}(c), we plot such frequencies where the transmission vanishes as a function of the wave vector for both $s$- and $p$-polarized incident light. Such frequencies vary quadratically as a function of $\mathbf{k}$, which is consistent with the band structure shown in Fig. \ref{phc_slab_design}(d). The transmission for both polarizations is very similar near $\mathbf{k} = 0$, confirming the polarization independence.

Figure \ref{transmission_plots}(d) plots the transmission spectra at normal incidence and at $|\mathbf{k}| = 0.05 \times 2\pi/a$ for both $s$ and $p$ polarizations. At $\mathbf{k} = 0$, the transmission spectra for the $s$ and $p$ polarizations are identical, as expected. At $|\mathbf{k}| = 0.05 \times 2\pi/a$, the transmission spectra for the two polarizations are also quite similar to each other, due to our design of the photonic crystal slab. For the differentiator, we choose the operation frequency to be $\omega_0=0.7902 \times 2\pi c/a$ where the transmission vanishes at $\mathbf{k} = 0$. For free-space compression, we note that the transmission is near unity at the frequency $\omega - \omega_0 = 1.58 \times 10^{-2}$  $2\pi c/a$ for both $\mathbf{k} = 0$, as well as for $|\mathbf{k}| = 0.05 \times 2\pi/a$. The transmission is in fact near unity at this frequency for all wave vectors between $0$ and $0.05 \times 2\pi/a$. We therefore choose this frequency $\omega$ as the operating frequency for free-space compression, since it requires a near-unity transmission coefficient for all wave vectors. In the numerical simulations below, we also choose the beam size such that the wave vector components of the beam are within the wave vector range of $0$ and $0.05 \times 2\pi/a$, corresponding to a numerical aperture of 0.062 and maximum incident angle of $\theta \approx 3.56 \degree$.  

We note that in our design, the same structure can perform two different functionalities by simply operating at different frequencies. Thus, the structure operates as a multifunctional metasurface.  
Previous works on multifunctional metasurfaces include multiwavelength achromatic metalenses \cite{capasso_metasurf_multiwavelength_nano_2018}, electro-optically controlled beam steering and focusing \cite{electro-optic_tunable_metasurf_atwater_2020}, and reprogrammable holograms \cite{reprog_hologram_metasurf_nature_2017}.

\section{Results}

\subsection{Free-space compression}

\begin{figure}[h]
\includegraphics[width=\linewidth]{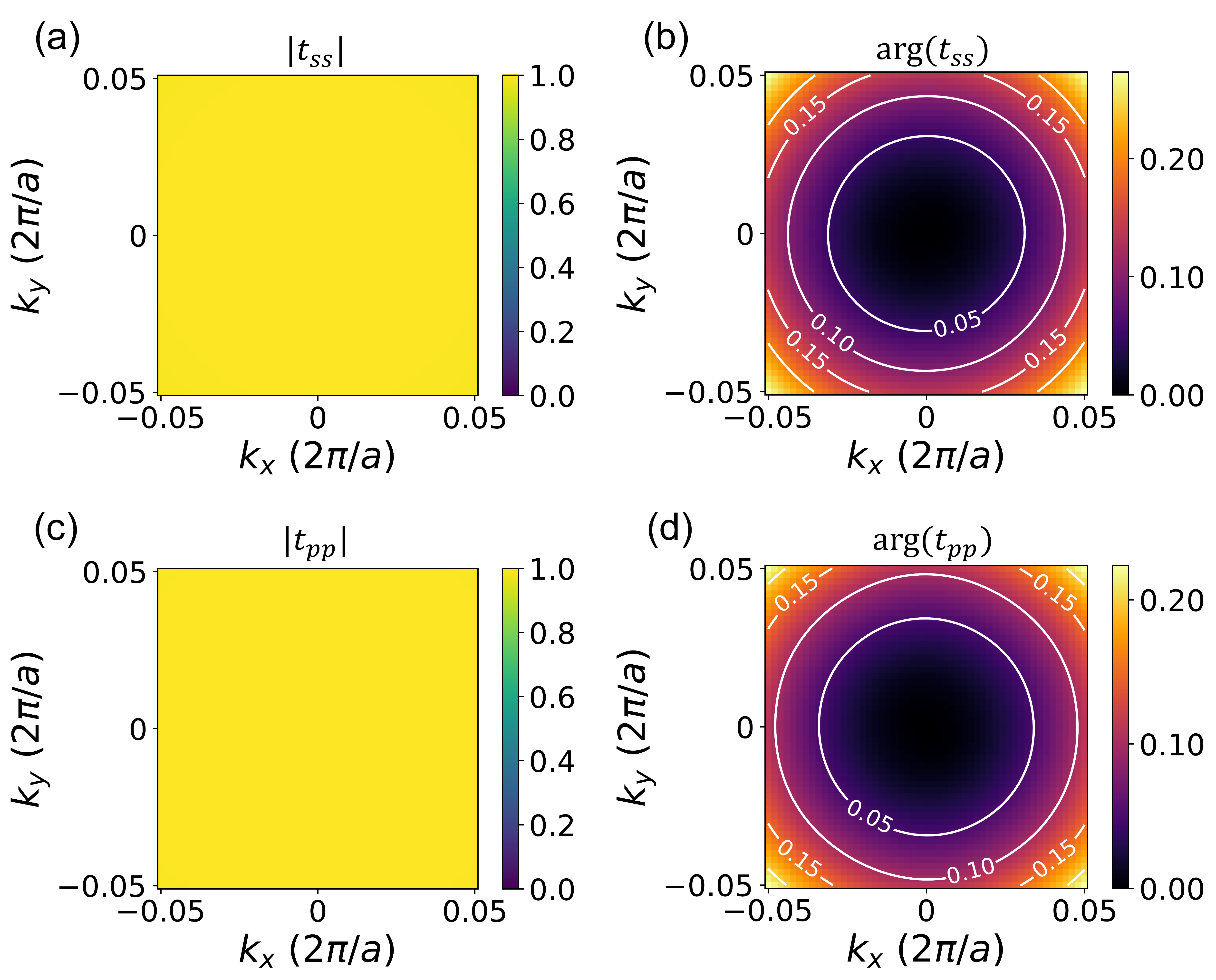}
\caption{\label{space_squeeze_trans_func} Transfer function for the free-space compressor at operating frequency $\omega_{op} = 0.806 \times 2\pi c/a$. (a) Calculated magnitude and (b) phase (rad) of $t_{ss}$ as a function of the in-plane wave vector $\mathbf{k} = (k_x, k_y)$, where $k_x, k_y \in [-0.05, 0.05] \times 2\pi/a$. (c) Calculated magnitude and (d) phase of $t_{pp}$ as a function of the in-plane wave vector $\mathbf{k}$.}
\end{figure}

Based on the discussion in the previous section, we now consider in more details the operation of the photonic crystal slab shown in Fig. \ref{phc_slab_design}(a) for both free-space compression and spatial differentiation. For free-space compression, at the operating frequency $\omega_{\text{op}} = 0.806 \times 2 \pi c/a$,
%
%
%
the transfer function as a function of the in-plane wave vector $\mathbf{k} $ is shown for both $s$ and $p$ polarizations in Figure \ref{space_squeeze_trans_func}. 
It was noted in Ref. \citenum{Guo_phc_diff_optica_2018} that for a photonic crystal slab having an isotropic band structure for its guided resonance, the $s$ and $p$ polarizations each excite only one of the two bands, and there is no polarization conversion induced by the guided resonance. 
Thus, in Figure \ref{space_squeeze_trans_func}, we plot the magnitude and phase for only the polarization-preserving transmission coefficients $t_{ss}$ and $t_{pp}$ for $s$- and $p$-polarized light, respectively. 
%
We observe that the transmission coefficient exhibits unity magnitude over the wave vector range of operation for both the $s$ (Fig. \ref{space_squeeze_trans_func}(a)) and $p$ (Fig. \ref{space_squeeze_trans_func}(b)) polarizations. Both the phases of $t_{ss}$ and $t_{pp}$ show an isotropic dependency on the direction of $\mathbf{k}$, as shown in Figures \ref{space_squeeze_trans_func}(b) and \ref{space_squeeze_trans_func}(d), respectively.

\begin{figure}[b]
\includegraphics[width=\linewidth]{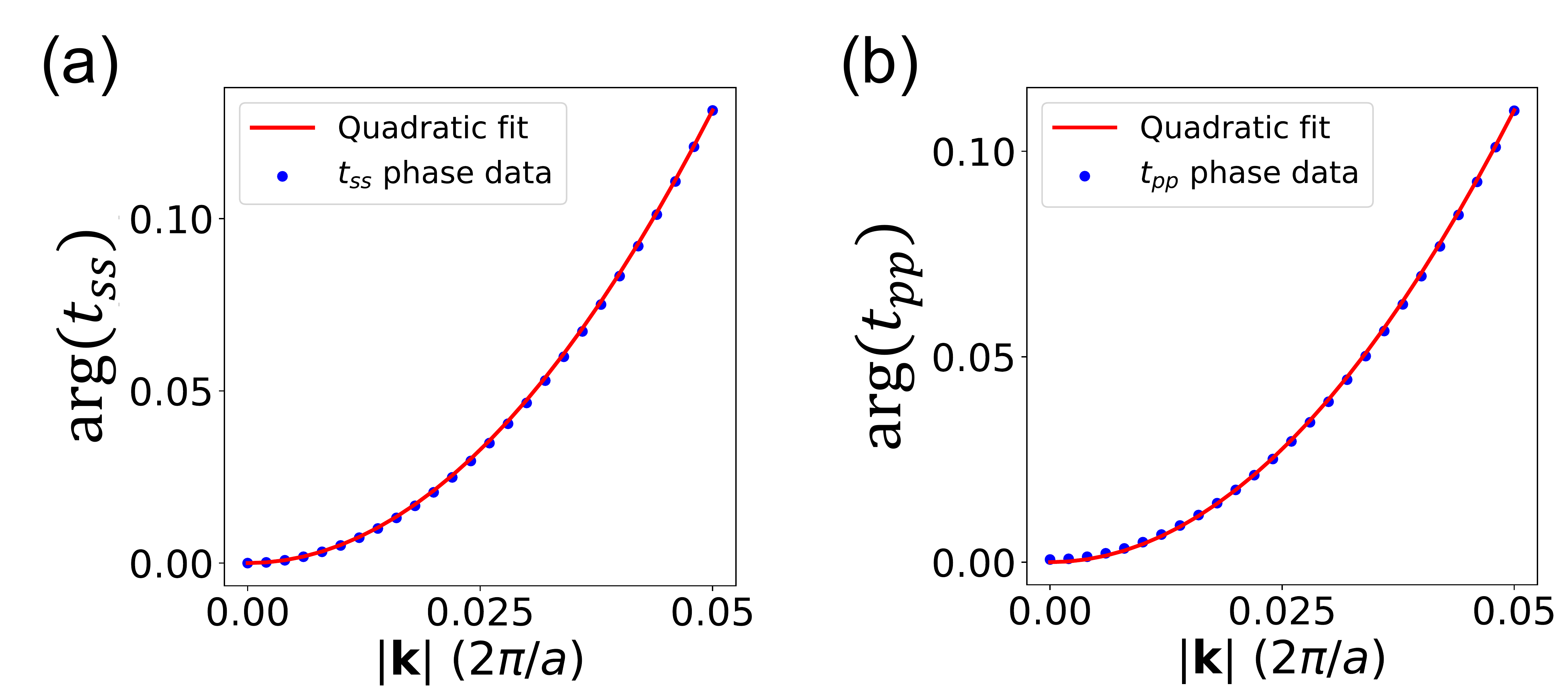}
\caption{\label{space_squeeze_quad_fits} Quadratic fits of (a) $t_{ss}$ phase and (b) $t_{pp}$ phase as a function of in-plane wave vector $\mathbf{k}$ at the operating frequency $\omega_{op} = 0.806 \times 2\pi c/a$.}
\end{figure}

The phases of $t_{ss}$ and $t_{pp}$ have quadratic dependence on the magnitude of $\mathbf{k}$, as confirmed in Figures \ref{space_squeeze_quad_fits}(a) and \ref{space_squeeze_quad_fits}(b), respectively.
The excellent quadratic fits of the phases confirm that Condition 2 for free-space compression is indeed satisfied.
The fitted functions are given by $\text{arg}(t_{ss}) = 1.332 * |\mathbf{k}|^2$ for the $s$ polarization and $\text{arg}(t_{pp}) = 1.114 * |\mathbf{k}|^2$ for the $p$ polarization, where the quadratic coefficients are expressed in units of $a^2$. Using the fitted quadratic coefficients, we can extract the effective distance $d_{\text{eff}}$ for each polarization (see Eqn. 20 in Ref. \citenum{Guo_squeeze_optica_2020}). 
We find that $d_{\text{eff}} = 13.49 a$ for $s$-polarized light and $d_{\text{eff}} = 11.29 a$ for $p$-polarized light. This shows that the different polarizations have very comparable rates of space compression, in contrast to previous works \cite{Guo_squeeze_optica_2020}. 
Since the thickness of our device is $d = 0.32a$, this yields compression ratios of approximately $ 42.17 $ and $ 35.27 $ for the $s$ and $p$ polarizations, respectively.

\begin{figure*}[t] 
\includegraphics[width=\linewidth]{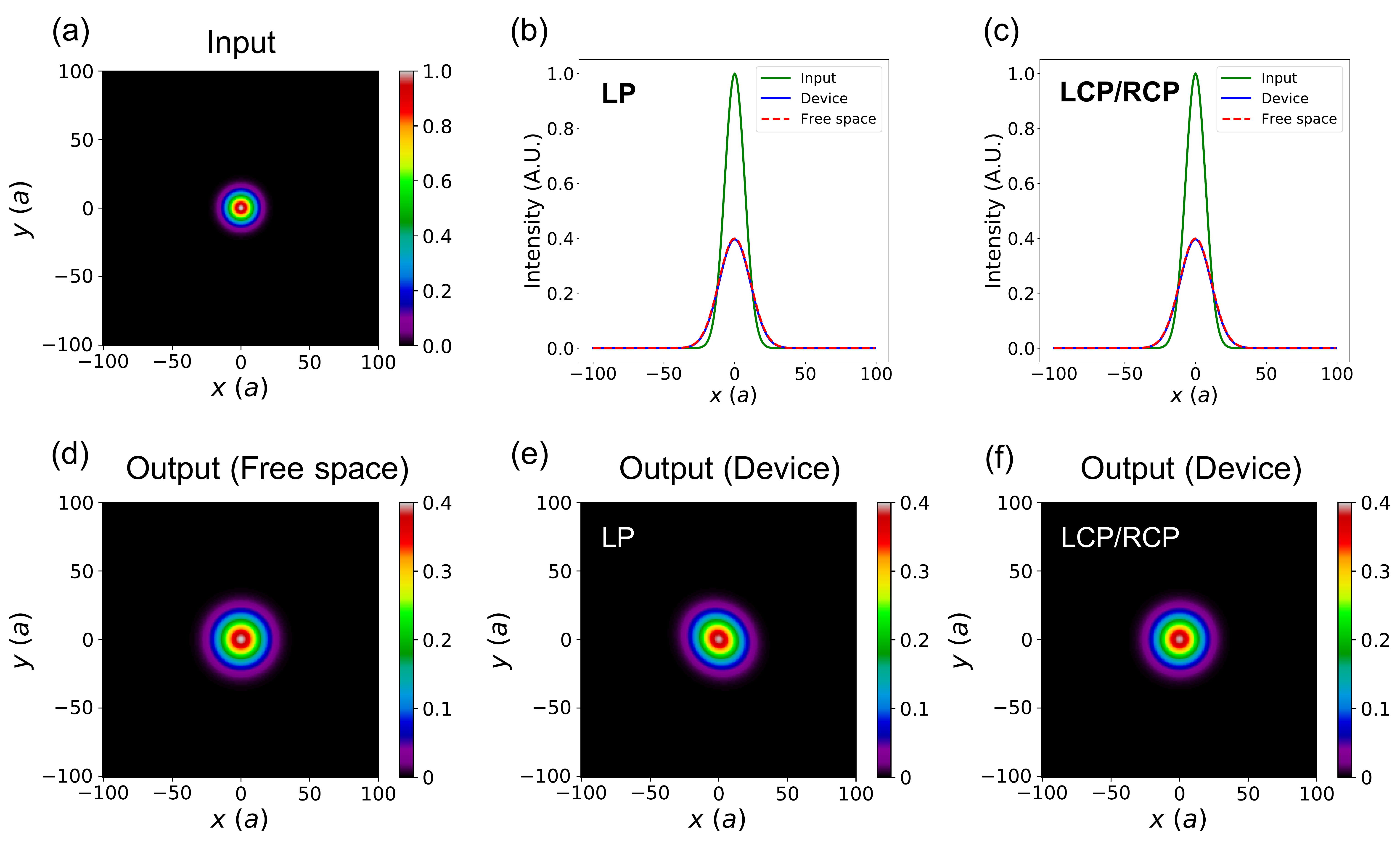}
\caption{\label{space_squeezer_num_demo}Numerical demonstration of free-space compressor at the operating frequency $\omega_{op} = 0.806 \times 2\pi c/a$. (a) Intensity spatial profile of input Gaussian beam that is linearly polarized (LP) or circularly-polarized (LCP/RCP). (b) Radial intensity profile of the input LP Gaussian beam, the output beam of 50 cascaded devices, and the output beam after the equivalent free-space propagation along $y=0$. (c) Radial intensity profile of the input LCP/RCP Gaussian beam, the output beam of 50 cascaded devices, and the output beam after the equivalent free-space propagation along $y=0$. (d) Intensity spatial profile of output beam after transmission through the equivalent free space for both polarizations. (e) Intensity spatial profile of the LP Gaussian beam and (f) the LCP/RCP Gaussian beam after 50 cascaded devices.}
\end{figure*}

We numerically demonstrate our space-compression device on an input Gaussian beam in Figure \ref{space_squeezer_num_demo}. The intensity profile of our input beam is shown in Figure \ref{space_squeezer_num_demo}(a). 
%
%
%
To generate the output beam profiles, we first Fourier transform the input Gaussian beam. Then, we multiply element-wise by the transfer function $t(\mathbf{k})$, which is calculated by decomposing the input polarization (given in the $x,y$ basis) into its $s$ and $p$ components, summing the contributions of $t_{ss}$ and $t_{pp}$ at each wave vector $\mathbf{k}$ in the operation range, then projecting the result back into the direction of the input polarization, expressed in the $x,y$ basis. Finally, we apply the inverse Fourier transform to obtain the output spatial field profile.
%
In our numerical demonstration, we use left circularly-polarized (LCP) and right circularly-polarized (RCP) light, as well as diagonal, linearly-polarized (LP) light as our input polarizations. To simulate a cascade of 50 devices, we apply the transfer function 50 times to the Fourier transform of our input Gaussian.

For the purpose of comparison, we plot the output intensity profile of the equivalent free-space propagation over a distance of $50d_\text{eff}\approx 621 a$ for LP and LCP/RCP light in Figure \ref{space_squeezer_num_demo}(d). 
The free-space output intensity plot is the same for each polarization. 
Compared with the input profile, the output profile is more extended in space due to diffraction.
We plot the output intensity profile of left and right circularly-polarized light in Figure \ref{space_squeezer_num_demo}(f), as well as its radial intensity profile in Figure \ref{space_squeezer_num_demo}(c). 
The beam width of free-space propagation is faithfully reproduced by our cascade of devices. In addition, the output beam shape matches that of free space.
Figure \ref{space_squeezer_num_demo}(e) shows the output intensity profile of diagonal, linearly-polarized light. 
Like LCP/RCP light, the beam width of the Gaussian is accurately reproduced for the linear polarization, as shown in the radial intensity profile plot (Figure \ref{space_squeezer_num_demo}(b)). 
The slight distortion from the ideal Gaussian beam shape is due to the slight difference in magnitude of the transfer functions for the $s$ and $p$ polarizations, the effect of which becomes more pronounced as more devices are cascaded. 
To mitigate the distortion effects, the structure can be further tuned to achieve better agreement between the transfer functions of the $s$ and $p$ polarizations. 
Note that we are able to employ a Gaussian beam in our demonstration due to the similar phase responses of both $s$ and $p$ polarizations. In contrast, 
the operations of the previous photonic crystal slab design  
are limited to cylindrical vector input beams \cite{Guo_squeeze_optica_2020}, since there must be a node at the origin of a purely $s$- or $p$-polarized beam due to the continuity of the electric field \cite{Zhan_cylind_vector_beams_2009}. 

\subsection{Second-order differentiation}

\begin{figure}[h]
\includegraphics[width=\linewidth]{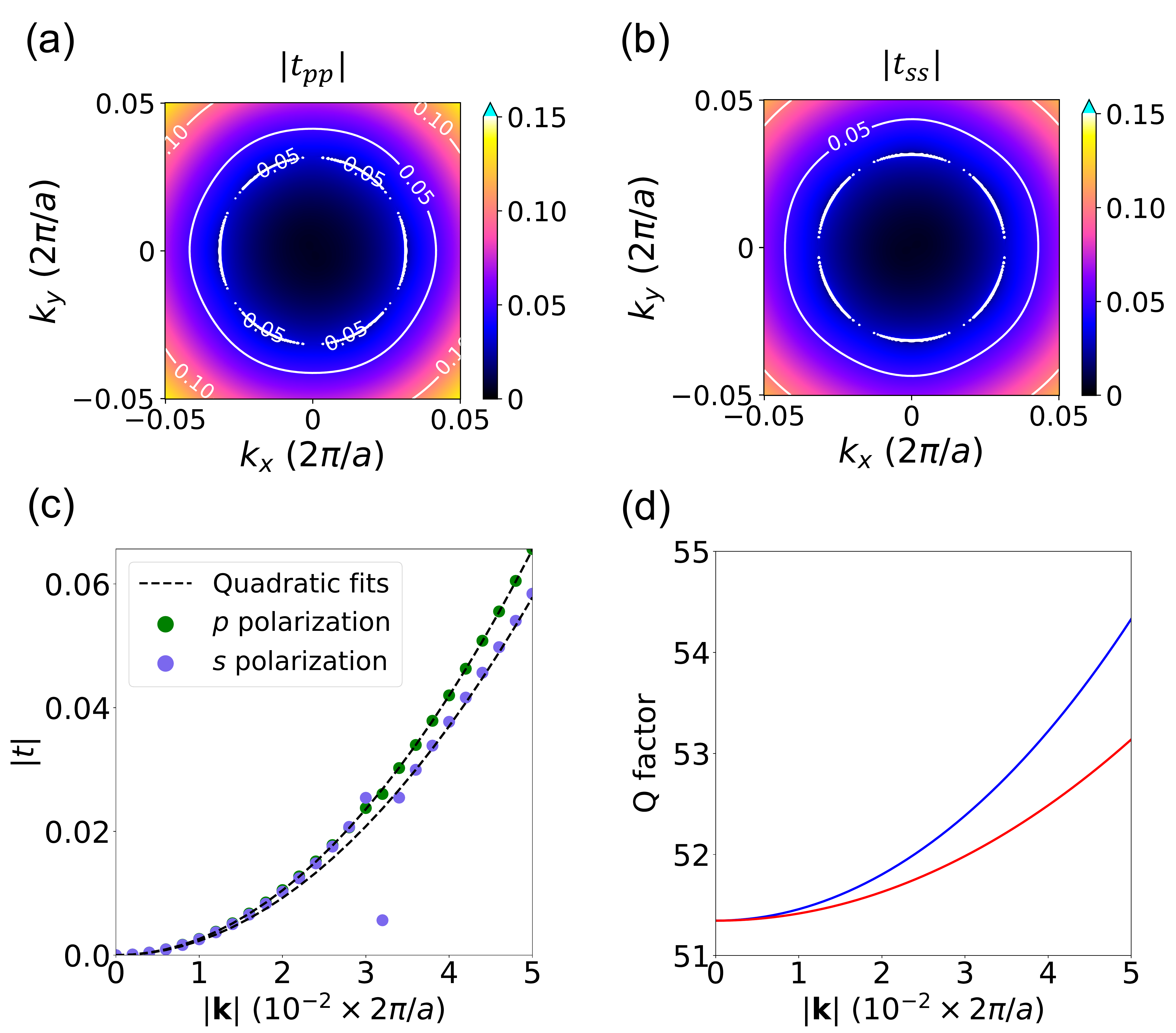}
\caption{\label{diff_trans_funcs} Transfer function for the differentiator at operating frequency $\omega_{op} = 0.7902 \times 2\pi c/a$. (a) Transmission coefficient $|t_{pp}|$ and (b) $|t_{ss}|$ computed at $\omega_{op}$ using RCWA. (c) Transmission coefficients $|t_{pp}|$ and $|t_{ss}|$ plotted as functions of the in-plane wavevector along the $\phi = 34 \degree$ direction at $\omega_{op}$. Quadratic fits are indicated by the dashed black lines. (d) Q factors of the two degenerate modes 
plotted as a function of the in-plane wave vector in the $\Gamma K$ direction, calculated using GME.}
\end{figure}

For our spatial differentiation functionality, we plot the transfer functions of each polarization at the operating frequency $\omega_{\text{op}}= 0.7902 \times 2\pi c/a$ in Figure \ref{diff_trans_funcs}.
Similar to the space compressor, the numerical aperture is 0.063, corresponding to a maximum incident angle of $\theta \approx 3.61 \degree$. 
%
%
%
The magnitudes of $t_{pp}$ and  $t_{ss}$ are shown in Figures \ref{diff_trans_funcs}(a) and \ref{diff_trans_funcs}(b), respectively. In both plots, we see that the transfer function is isotropic, in agreement with the band structure. Note that the discontinuities on the innermost contour are due to band crossings, which are also visible in the transmission plots (e.g. Fig. \ref{transmission_plots}(b) shows the band crossing in $t_{pp}$ along the $\Gamma K$ direction). However, these do not affect the performance of our differentiator, since point discontinuities only have a weak effect on 
the overall functional form of the transfer function in terms of the in-plane wave vector $\mathbf{k}$. 

In Figure \ref{diff_trans_funcs}(c), we confirm the relation $|t(\mathbf{k})| \propto |\mathbf{k}|^2$ by fitting to a quadratic function, as required by Condition 2. 
The quadratic fits are given by $|t_{ss}| = 23.13*|\mathbf{k}|^2$ and $|t_{pp}| = 26.21*|\mathbf{k}|^2$ for the $s$ and $p$ polarizations, respectively. This slight difference in curvature between the polarizations is a result of the different Q factors of the two degenerate guided modes, which are plotted in Figure \ref{diff_trans_funcs}(d). At $\mathbf{k} = 0$, the calculated Q factor for both modes is $51.35$, and the difference increases away from $\mathbf{k} = 0$. 
This difference is also visible from the transmission plots of Figure \ref{transmission_plots}, where by comparing Fig. \ref{transmission_plots}(b) with Fig. \ref{transmission_plots}(a), 
we observe that the guided resonance for the $p$ polarization has a more narrow radiative linewidth away from $\mathbf{k} = 0$ than that for the $s$ polarization.

\begin{figure}[h]
\includegraphics[width=\linewidth]{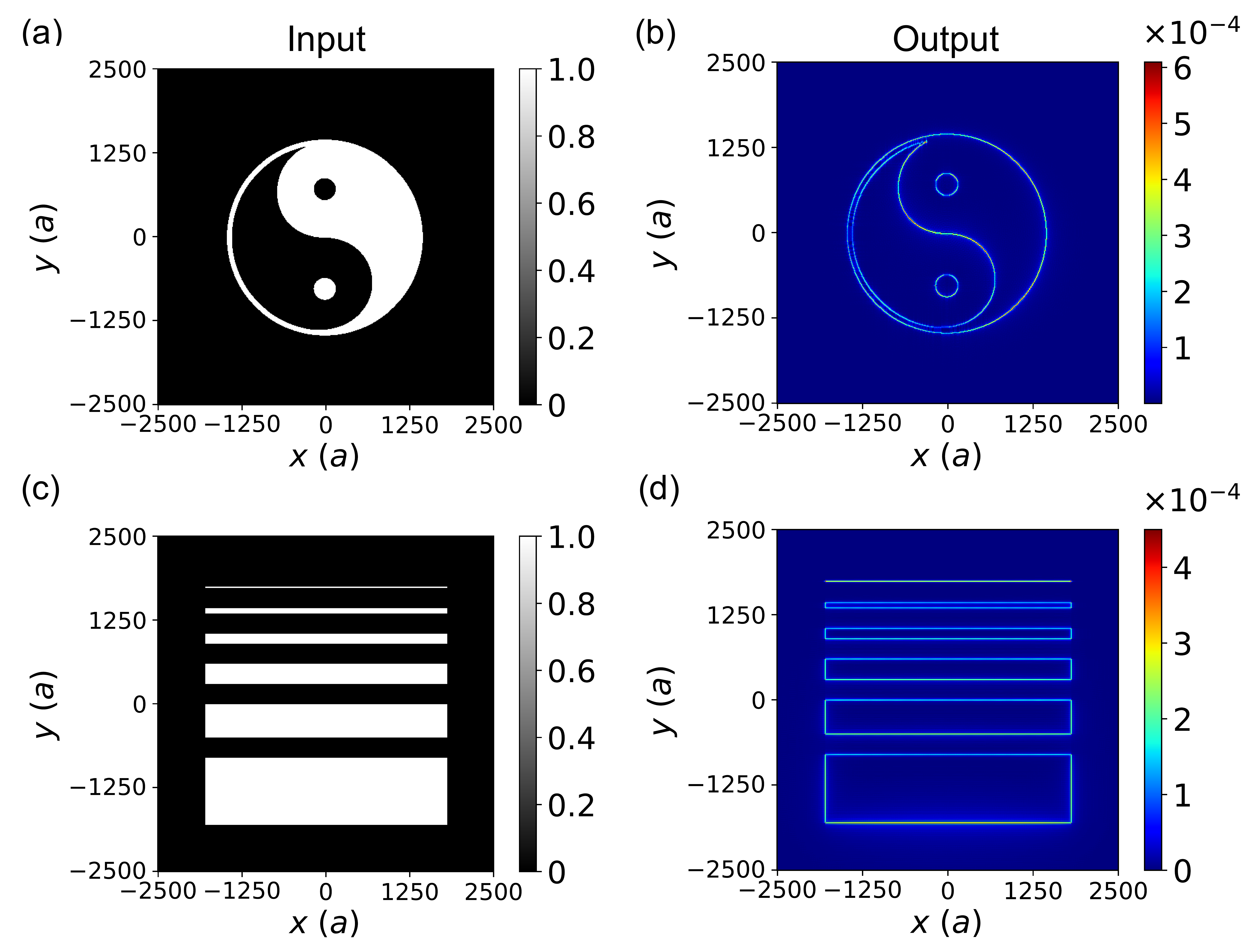}
\caption{\label{diff_num_demo} Numerical demonstration of the differentiator at the operating frequency $\omega_{op} = 0.7902 \times 2\pi c/a$. 
(a) Intensity spatial profile of incident image of yin and yang using left circularly-polarized light. (b) Output intensity of transmitted image of yin and yang. (c) Intensity spatial profile of incident image of slot patterns using left circularly-polarized light. (d) Output intensity of transmitted image of slot patterns. }
\end{figure}

We numerically demonstrate our isotropic, second-order differentiator in Figure \ref{diff_num_demo}. To generate the output field profile, we first Fourier transform the input image, multiply element-wise by the transfer function $t(\mathbf{k})$ for the input light's polarization, then apply the inverse Fourier transform. In our demonstration, we employ LCP incident light. The input image of the yin and yang symbol is shown in Figure \ref{diff_num_demo}(a), while the differentiated output is depicted in Figure \ref{diff_num_demo}(b). 
The detection of edges along each direction demonstrates the isotropic differentiation of our device. We further test the spatial resolution of our device using the input image shown in Figure \ref{diff_num_demo}(c). The slot patterns have widths of $1000a$, $500a$, $300a$, $150a$, $80a$, and $20a$. From the output intensity profile in Figure \ref{diff_num_demo}(d), we see that the resolution limit is about $20a$. Using RCP incident light gives an identical response (not shown), demonstrating polarization-independent operation. 

\section{Discussion and Conclusion}

In summary, we have designed a photonic crystal slab that can nonlocally perform polarization-independent, isotropic differentiation and free-space compression when operated at different frequencies at normal incidence. By introducing daisy-shaped air holes, 
we can achieve guided resonances with band curvatures that are near-degenerate for the two polarizations near $\mathbf{k} = 0$. Note that only the real parts of the band dispersions were near-degenerate, while the imaginary parts or the quality factors of the modes differed. 
Future directions include optimizing the Q factors to be degenerate as well to achieve polarization-independence over a wider wave vector range. 
%
%
This work demonstrates the potential of using dispersion engineering in photonic crystal slabs to achieve desired polarization-independent functionalities in ultrathin devices. 

\begin{acknowledgments}
The authors thank Haiwen Wang for helpful discussion. This work is supported by MURI grants from the U. S. Air Force Office of Scientific Research (Grant Nos FA9550-21-1-0312, and FA9550-17-1-0002), and used the Extreme Science and Engineering Discovery Environment (XSEDE), which is supported by National Science Foundation grant number ACI-1548562.

O.L. acknowledges support from the NSF Graduate Research Fellowship and the Stanford Graduate Fellowship.
\end{acknowledgments}



\bibliography{apssamp}

\end{document}